\documentclass[sigconf]{acmart}
\usepackage{booktabs} 
\usepackage[nolist]{acronym}
  \newacro{ml}[ML]{Machine Learning}
  \newacro{ai}[AI]{Artificial Intelligence}
  \newacro{dl}[DL]{Deep Learning}
  \newacro{mlp}[MLP]{Multi-Layer Perceptron}
  \newacro{cnn}[CNN]{Convolutional Neural Network}
  \newacro{dnn}[DNN]{Deep Neural Network}
  \newacro{bnn}[BNN]{Binary Neural Network}
  \newacro{fm}[FM]{Feature Map}
  \newacro{relu}[ReLU]{Rectified Linear Unit}
  \newacro{alm}[ALM]{Adaptive Logic Module}
  \newacro{mnist}[MNIST]{Modified National Institute of Standards and Technology}
  \newacro{ilsvrc}[ILSVRC]{ImageNet Large Scale Visual Recognition Competition}
  \newacro{asic}[ASIC]{Application Specific Integrated Circuits}
  \newacro{dram}[DRAM]{Dynamic Random Access Memory}
  \newacro{fpga}[FPGA]{Field Programmable Gate Array}
  \newacro{gpu}[GPU]{Graphics Processing Unit}
  \newacro{gpp}[GPP]{General Purpose Processor}
  \newacro{cpu}[CPU]{Central Processing Unit}
  \newacro{pe}[PE]{Processing Element}
  \newacro{simd}[SIMD]{Single Instruction on Multiple Data}
  \newacro{simt}[SIMD]{Single Instruction on Multiple Threads}  
  \newacro{gemm}[GEMM]{General Matrix Multiplication}
  \newacro{ewmm}[EWMM]{Element-Wise Matrix Multiplication}
  \newacro{fft}[FFT]{Fast Fourier Transofrm}
  \newacro{dsp}[DSP]{Digital Signal Processing}
  \newacro{hdl}[HDL]{Hardware Description Language}
  \newacro{le}[LE]{Logic Elements}
  \newacro{fifo}[FIFO]{First-In First-Out}
  \newacro{haddoc}[HADDOC]{Hardware Automated Dataflow Description Of CNNs}
  \newacro{hls}[HLS]{High-Level Synthesis}
  \newacro{hpc}[HPC]{High Performance Computing}  
  \newacro{moc}[MoC]{Model of Computation}
  \newacroplural{moc}[MoC]{Models of Computation}
  \newacro{ocr}[OCR]{Optical Character Recognition}
  \newacro{qos}[QoS]{Quality of service}
  \newacroplural{qos}[QoS]{Qualities of service}
  \newacro{tpr}[TPR]{True Positive Rate}
  \newacro{mac}[MAC]{Multiply Accumulate}
  \newacro{fc}[FC]{Fully Connected}
  \newacro{simd}[SIMD]{Single Instruction on Multiple Data}
  \newacro{vhdl}[VHDL]{VHSIC Hardware Description Language}
  \newacro{lut}[LUT]{Look-Up Table}
  \newacro{nef}[NEF]{Neighborhood Extraction Factorization}
  \newacro{ne}[NE]{Neighborhood Extraction}
  \newacro{lut}[LUT]{Lookup Table}
  \newacro{hdl}[HDL]{Hardware Description Language}
  \newacro{rtl}[RTL]{Register Transfer Level}
  \newacro{ip}[IP]{Intellectual Property}
  \newacro{dhm}[DHM]{Direct Hardware Mapping}
  \newacro{dag}[DAG]{Direct Acyclic Graph}
  \newacro{dpn}[DPN]{Data-flow Process Network}
  \newacro{sdf}[SDF]{Static Data-Flow}
  \newacro{sdfg}[SDFG]{Synchronous DataFlow Graph}
  \newacroplural{ip}[IP]{Intellectual Properties}
  \newacro{sfp}[SFP]{Static Fixed Point}
  \newacro{dfp}[DFP]{Dynamic Fixed Point}
  \newacro{ttq}[TTQ]{Trained Ternary Quantization}
  \newacro{mcm}[MCM]{Multiple Constant Multiplication}
  \newacro{scm}[SCM]{Single Constant Multiplication}
  \newacro{moa}[MOA]{Multiple Operand Adder}
  \newacro{loa}[LOA]{Lower-part-Or}
  \newacro{mred}[MRED]{Mean Relative Error Distance}

\usepackage{tikz}
\usepackage{multirow}

\copyrightyear{2018} 
\acmYear{2018} 
\setcopyright{acmcopyright}
\acmConference[SAMOS XVIII]{2018 International Conference on Embedded
Computer Systems: Architectures, Modeling, and Simulation}{July 15--19,
2018}{Pythagorion, Samos Island, Greece}
\acmBooktitle{SAMOS XVIII: 2018 International Conference on Embedded
Computer Systems: Architectures, Modeling, and Simulation, July 15--19,
2018, Pythagorion, Samos Island, Greece}
\acmPrice{15.00}
\acmDOI{10.1145/3229631.3235024}
\acmISBN{978-1-4503-6494-2/18/07}

\begin{document}
\title{The Challenge of Multi-Operand Adders in CNNs on FPGAs}
\subtitle{How Not to Solve It!}

\author{Kamel Abdelouahab}
\orcid{1234-5678-9012}
\affiliation{%
  \institution{Institut Pascal UMR CNRS 6602\\Universite Clermont Auvergne}
  \city{Clermont Ferrand}
  \state{France}
}

\author{Maxime Pelcat}
\orcid{1234-5678-9012}
\affiliation{%
  \institution{IETR UMR CNRS 6164, Institut Pascal \\INSA Rennes}
  \city{Rennes}
  \state{France}
}

\author{Francois Berry}
\orcid{1234-5678-9012}
\affiliation{%
  \institution{Institut Pascal UMR CNRS 6602\\Universite Clermont Auvergne}
  \city{Clermont Ferrand}
  \state{France}
}
\renewcommand{\shortauthors}{K. Abdelouahab et al.}

\begin{abstract}
Convolutional Neural Networks (CNNs) are computationally intensive algorithms that currently require dedicated hardware to be executed. In the case of FPGA-Based accelerators, we point-out in this work the challenge of Multi-Operand Adders (MOAs) and their high resource utilization in an FPGA implementation of a CNN. To address this challenge, two optimization strategies, that rely on time-multiplexing and approximate computing, are investigated. At first glance, the two strategies looked promising to reduce the footprint of a given architectural mapping, but when synthesized on the device, none of them gave the expected results. Experimental sections analyze the reasons of these unexpected results.

\end{abstract}

\maketitle

\section{Introduction}
Since their breakthrough in 2012, Deep \acp{cnn}~\cite{Krizhevsky2012a} have become the \textit{de-facto} standard used to solve an ever greater number of computer-vision tasks that range from image classification to semantic segmentation and scene recognition~\cite{He2016b,Long2015,Redmon2018}.  However, \ac{cnn}-based algorithms are computationally intensive and their execution in real-time remains a challenging task, especially in embedded devices.

To address this challenge, a variety of dedicated accelerators, built around \acp{fpga} and \acp{gpu}, have been proposed. A key advantage of the former solution is its superior power efficiency when compared to the latter~\cite{Nurvitadhi2017}. Moreover, \ac{cnn} workloads have a streaming nature that is well suited to reconfigurable hardware architectures such as \acp{fpga}, which motivated numerous research efforts to optimize \ac{fpga} implementation for \acp{cnn}~\cite{Zhang2015,Qiu2016,Ma2018}. Among the proposed methods, one possibility is to \textit{directly map} a \ac{cnn} graph on the \ac{fpga} resources,  allocating each processing actor its own hardware instance, and each edge of the graph its own \ac{fifo} channel~\cite{Abdelouahab2017}.

In this paper, we point-out to a key feature of this \ac{dhm}, which is the high hardware cost of \textit{Multi-Operand-Adders}. More particularly, we found that 69\% of the logic used to map the \ac{cnn} graph on an \ac{fpga} is allocated to logic implementing \emph{aggregated} adders which have the particularity to receive operands per thousands. To reduce these footprint of adders, we investigate in this work two strategies based on time-multiplexed serialization and approximate computing. Each method is promising on paper, but result in unexpectedly bad results when synthesized on FPGAs. Our experiments are reproducible and available on-line\footnote{\url{https://github.com/KamelAbdelouahab/Multi-Operand-Adder}}.

\section{Multi-Operand-Adders in CNNs}

A \ac{cnn} graph takes the form of a succession of layers that hierarchically extract features from raw inputs. Most computation occurs in the $convolution$ layers which rely on a learned set of $N$ three-dimensional convolution filters of size $C \times J \times K$ to output a 3D feature map of size $N \times V \times U$. Thus, each filter involves a dot-product of $C \times J \times K$ elements as shown in equation \ref{eq:convLayer}.

\begin{align}
\label{eq:convLayer}
\forall & \left\{n, u, v \right\} \in \left[ 1,N \right] \times \left[ 1,V \right] \times \left[ 1,U \right] \nonumber \\
    & {Y}[n,v,u] = \sum_{c=1}^{C^{}} \sum_{j=1}^{J}  \sum_{k=1}^{K} {X}[c,v+j,u+k] . {\Theta}[n,c,j,k]
\end{align}

\begin{figure}
\begin{tikzpicture}[scale=0.65, every node/.style={scale=0.72}]

    \tikzset{prod/.style={draw,rectangle,minimum width=1.5cm}}; 
    \tikzset{adder/.style={draw,circle,minimum size=0.5cm}}; 
    \tikzset{io/.style={}};
    
    \node [io] (p0) at (-0.2,0)  {$X_{000}$};
    \node [io] (p1) at (-0.2,-1) {$X_{001}$};
    \node [io] (pk) at (-0.2,-3) {$X_{CJK-2}$};
    \node [io] (pn) at (-0.2,-4) {$X_{CJK-1}$};

    
    \node [prod] (prod0) at (2,0)  {$\times \: \theta_{000}$};
    \node [prod] (prod1) at (2,-1) {$\times \: \theta_{001}$};
    \node [prod] (prodk) at (2,-3) {$\times \: \theta_{CJK-2}$};
    \node [prod] (prodn) at (2,-4) {$\times \: \theta_{CJK-1}$};
    \node [io]   (prodd) at (2,-1.8)   {$\vdots$};
    
    \node [adder] (add0) at (4,0) {$+$};
    \node [adder] (add1) at (4,-1) {$+$};
    \node [adder] (add2) at (4,-4) {$+$};
    \node [adder] (add3) at (4,-3) {$+$};  
    \node [adder] (add4) at (5,-0.5) {$+$};
    \node [adder] (add5) at (5,-3.5) {$+$};
    \node [adder] (add6) at (6,-1.5) {$+$};
    \node [adder] (add7) at (6,-2.5) {$+$};
    \node [adder] (add8) at (7,-2) {$+$};
    \node (y) at  (8.2,-2) {$Y$};
    
    \node (addp) at (5,-1.9) {$\hdots$};
    \node (addd) at (4,-1.8) {$\vdots$};
    \node (pd)   at (0,-1.8) {$\vdots$};

    \draw[->,>=latex] (p0)--(prod0);
    \draw[->,>=latex] (p1)--(prod1);
    \draw[->,>=latex] (pk)--(prodk);
    \draw[->,>=latex] (pn)--(prodn);
    
    
    \draw[->,>=latex] (prod0.east)--(add0);
    \draw[->,>=latex] (prod1.east)--(add0);
    \draw[->,>=latex] (prodk.east)--(add2);
    \draw[->,>=latex] (prodn.east)--(add2);
    
    \draw[->,>=latex] (add1)--(add4);
    \draw[->,>=latex] (add0)--(add4);
    \draw[->,>=latex] (add2)--(add5);
    \draw[->,>=latex] (add3)--(add5);
    \draw[->,>=latex] (add6)--(add8);
    \draw[->,>=latex] (add7)--(add8);
    \draw[->,>=latex] (add8)--(y);
    
    \draw[thick,dotted] (3.4,1.5) rectangle (7.5,-4.5);
    \draw[thick,dotted] (0.8,1.5) rectangle (3.2,-4.5);
    
    \node (moa_text) at (5.7,1.2) {Multi-Operand Adder};
    \node (comment) at (5.8,0.7) {(Binary Adder tree)};
    
    \node (scm_text) at (2.1,1.2) {Constant};
    \node (comment) at (2.1,0.7) {Multipliers};

\end{tikzpicture}
\caption{Direct Hardware Mapping of Dot Products in convolution layers: A Binary Adder tree sums the partial-products}
\label{scm_dp}
\end{figure}
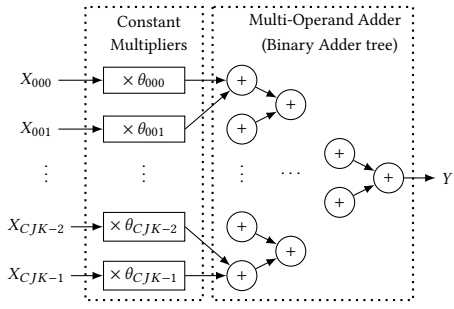

A method to accelerate the execution of these layers is to fully unroll the parallel computations involved in dot-products, and to map each multiplication to a dedicated hardware instance, as illustrated in Figure~\ref{scm_dp}. In \acp{fpga}, the advantage of this \ac{dhm} strategy is to tile the circuitry of the multiplier according to the value of the multiplicand (i.e convolution filter) by applying \ac{scm} optimization techniques\cite{Voronenko2007a} where, for instance, multiplications by zero are removed and multiplications by a power of two are implemented by shift registers. As an example, a \ac{dhm}-based implementation of the LeNet5 network requires $\times 8.6$ less logic elements with this \ac{scm} optimization than without it, as detailed in~\cite{Abdelouahab2017}.

A drawback of the DHM solution is that each layer requires $N$ \acp{moa} with $C \times J \times K$ inputs in order to accumulate the partial products. By default, synthesis tools instantiate deep binary adder trees\footnote{Binary adders refer to adders with TWO operands and NOT adders with a 1-bit operand} to implement \acp{moa}, which require $CJK-1$ binary adders.
However, for state-of-the-art \acp{cnn}, C,J,K can be large, leading to adders with up to 1774 operands (cf table~\ref{adders-alexnet}.) As a result, most of the logic required to map a CNN layer is dedicated to the \acp{moa} part, which corresponds, for instance, to 69\% of the resources in the first layer of an AlexNet.

\begin{table}[h]
\centering
\caption{Number of MOAs and number of mean non-null inputs per Adder in AlexNet layers}
\label{adders-alexnet}
\begin{tabular}{|c|c|c|c|c|c|}
\hline
Layer     & conv1 & conv2 & conv3 & conv4 & conv5 \\ \hline
$N$       & 96    & 256   & 384   & 384   & 256   \\ \hline
$n_{opd}$ & 325   & 957   & 1774  & 1398  & 1420  \\ \hline
\end{tabular}
\end{table}

\section{Explored Solutions}
In order to reduce the hardware resources instantiated during the mapping of a given convolutional layer, we investigate two strategies that are commonly used to reduce the footprint of \acp{moa}. The first method iterates the accumulation of partial-sums through multiple clock cycles, leading to serialized adders. The second method relies on approximate computing techniques. 

\subsection{Serializing a cluster of adders}
\begin{figure}[!h]
\includegraphics[width=0.5\columnwidth]{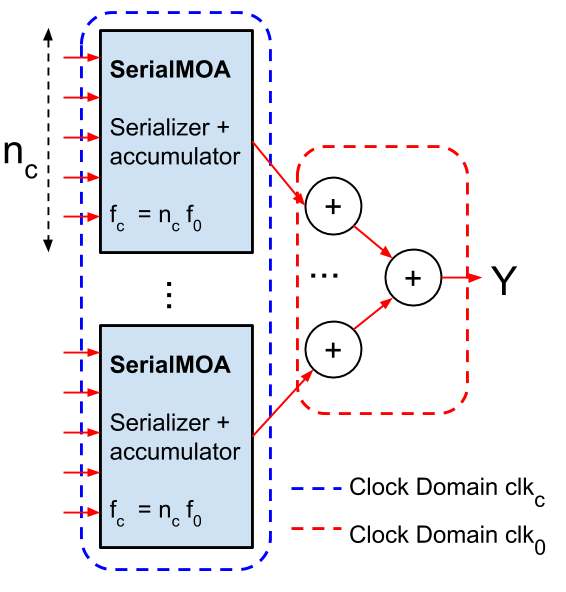}
\includegraphics[width=0.35\columnwidth]{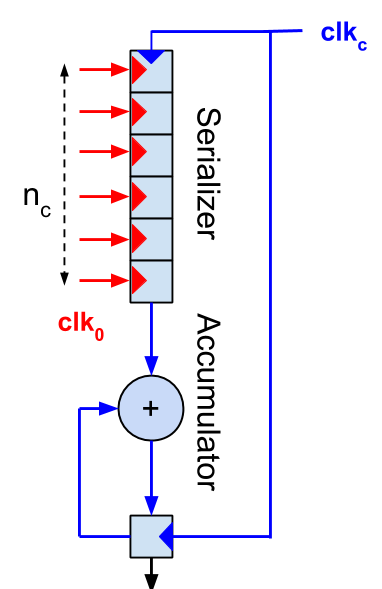}
\caption{Architecture of a serial MOA. Each Serializer Accumulator Pair replaces a cluster of adders in the MOA of Figure~\ref{scm_dp}}
\label{SerialMOA}
\end{figure}

FPGA devices --and more particularly the \ac{dsp} blocks they embed-- can run at a peak frequency that is much higher than the rate at which data and feature maps are acquired by a given \ac{cnn} layer ($\sim200$ MHz for a DSP Block versus about $27.6$ MHz for a 720p video stream). Given this, one can replace a cluster of binary adder trees by a serial accumulator that runs in a different, higher clock domain. In other words, we trade a clusters of $n_c$ binary adders that previously operated at a frequency $f_0$ for a \textit{single accumulator} that operates at a frequency $f_c$. In this context, $f_c = n_c f_0$ where $ 0 \leq n_c \leq n_{opd}$ and $n_{opd}$ is the number of adder operands. In return, a parallel-to-serial register (serializer) is required to input the accumulator, as shown in Figure~\ref{SerialMOA}. In recent \ac{fpga} devices, this method can replace an $n_c \approx 6$-input \ac{moa} by a single accumulator and a pair of serializers, which may reduce the footprint of the \ac{moa} by a factor of $n_c - 1 \approx 5$ under the hypothesis that serializers have a simpler circuitry when compared to \acp{moa}.

\subsection{Approximate Adders}
Deep \acp{cnn} are over-parametrized networks that tolerate by nature a degree of approximate computing. Approximations especially make sense during the inference phase because there is no error accumulation. Multiple state-of-the-art publications demonstrate the resiliency of \acp{cnn} towards compact bit-width arithmetic\cite{Suyog2015,Wu2016} and even binarization\cite{Courbariaux2016a,Rastegari2017}, which hints that \acp{cnn} may support others types of approximate computing techniques such as approximate adders. 
These adders, which use is limited to fault-tolerant applications, are known to deliver higher speed and power efficiency than exact operators~\cite{Jiang2015}. 

\begin{figure}[!h]
\includegraphics[width=0.7\columnwidth]{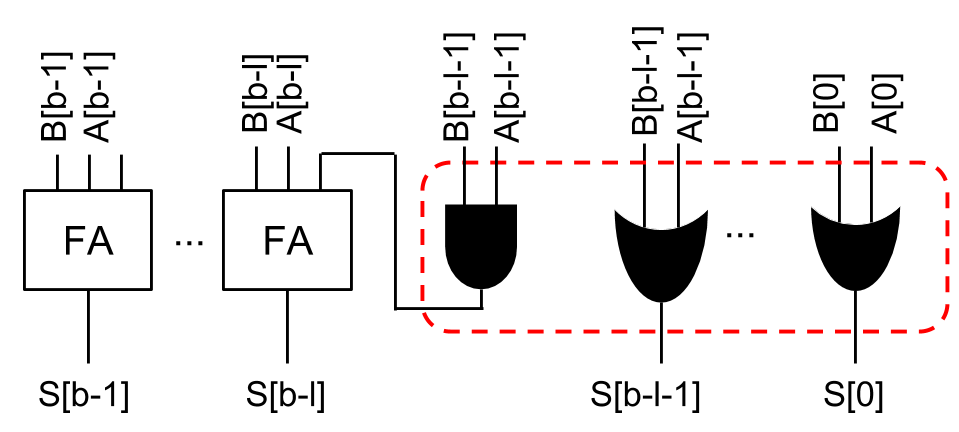}
\caption{Hardware structure of a Lower-part OR approximate adder (LOA). Approximate parts in the red box. Each LOA Replaces a Binary Adder in the Tree of figure ~\ref{scm_dp}}
\label{LOA}
\end{figure}

In order to solve the challenge of \ac{moa} footprint reduction for \acp{cnn}, we leverage on the low resource utilization of the \ac{loa} approximate adders~\cite{Mahdiani2010}. An \ac{loa} divides a $b$-bit adder into two sub-adders. The first one is an approximate $l$-bit sub-adder that computes the sum of least-significant bits by using a bit-wise OR operation. The second is an exact $(b-l)$-bit sub-adder that processes the most-significant bits using full adders. An extra AND gate is used to generate the carry-in signal for the exact adder part, as illustrated in Figure~\ref{LOA}. 

As pointed-out in the study of~\cite{Jiang2015}, \ac{loa} is the slowest but the most area efficient approximate adder, making it the best candidate for our study. In the Multi-Operand case, area saving may be achieved by replacing the exact binary adders in the tree with approximate adders such the \acp{loa}. 

\section{Experiments and Negative Results}
\subsection{Serialization}
\label{serial_exp}
\begin{figure}[!h]
\includegraphics[width=0.75\columnwidth]{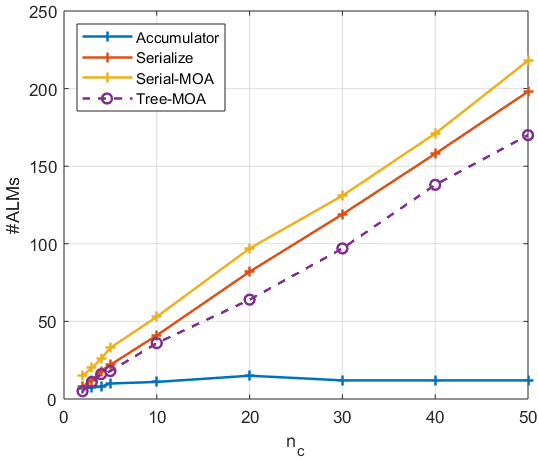}
\caption{Comparison of the Logic resources used by a serialized and fully pipelined implementation of a MOA: The serializer results in a large resource overhead}
\label{Jnc}
\end{figure}

In order to study the impact of serialization on an \ac{moa}, we design\footnote{Circuits are described in VHDL and synthesized on an Intel Stratix V 5SGXEA7 FPGA using Quartus 16.0. The bit-width of operands is 8 bits} and synthesize the Serializer/accumulator pair of Figure~\ref{SerialMOA}. Figure~\ref{Jnc} reports the logic utilization (in terms of \acp{alm}) of both the serializer, the accumulator and the serial adder for variable cluster sizes. These results are compared to the logic utilization of the standard binary adder tree implementation of a \ac{moa} (in dashed line).

This figure shows a very unexpected result. The resources utilization of the serializer/accumulator pair exceeds the resources used by a fully pipelined implementation of an \ac{moa} (i.e a binary adder trees). This is the result of the costly logic fabric required by the serializer part, displayed in Figure~\ref{Jnc}, which grows linearly with the number of parallel inputs (i.e operands). The overhead of serializers thus invalidate the approach.

\subsection{Approximate Adders}

In order to study the approximate \ac{loa} adder, we observe the impact of the \textit{approximation ratio} on both the accuracy and hardware utilization of a binary adder. The approximation ratio is defined as the number of approximated bits per total bit-width $l/b$. A ratio of 0\% corresponds to an exact adder while a ratio of 50\% means that half of the bits of a given addition have been approximatively processed using OR gates. 

To evaluate the accuracy of the method, the \ac{mred} metric is used. Let $s=x+y$ be the result of an exact addition of $x$ and $y$, and $\hat{s}$ the result of an approximate addition with same operands. The error distance is defined as:

\begin{equation}
MRED(s,\hat{s}) = mean \left( \frac{|\hat{s}-s|}{s}\right).
\end{equation}

\begin{figure}
\includegraphics[width=\columnwidth]{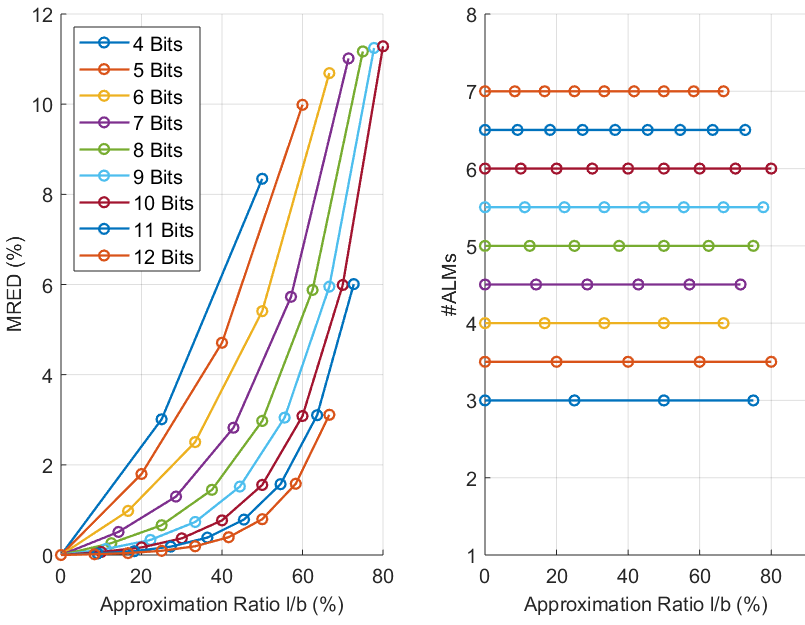}
\caption{Error Rates and logic utilization of \acp{loa} for variable bit-widths and approximation ratios.}
\label{loa_res}
\end{figure}

The evolution of the \ac{mred} metric when varying bit-widths and approximation ratios is illustrated in Figure~\ref{loa_res}, as well as their corresponding logic utilization. 

In terms of accuracy, using lower-part OR Adders results in a relatively small error ($< 10\%$ MRED for 8bits adders), which suggests that they might be exploited to derive energy-efficient \ac{cnn} accelerators. However, in terms of hardware utilization, our experiments show that no area saving can be achieved on an \ac{fpga} when using \acp{loa}. Indeed, the number of ALMs remains surprisingly constant, independently from the number of bits processed by an OR gate. This is explained by the fact that modern \ac{fpga} devices embed complex logical modules (ALM at Intel, Logical Blocks at Xilinx) that already contain a hard-wired full adder. This logical module either implements a full adder in the case of exact adders, or implements an OR gate in the case of approximate \ac{loa} adder. As a consequence, current \ac{fpga} and related hardware synthesizers do not benefit from approximate computing when targeting \ac{moa} adders and these results have been observed on both Intel and Xilinx \acp{fpga}.


\section{Conclusion}
This paper has introduced the challenge of multi-operand adder footprint reduction when implementing a \acl{cnn} with direct hardware mapping on an \ac{fpga}. Two potential solutions have been studied, relying on serialization of adders and approximate computing.
Though originally promising, these solutions have proven ineffective with current \ac{fpga} architectures that do not lend themselves well to adder approximation and serialization. The serialization of a cluster of adders does not reduce the footprint since the serializers require too many logic elements. The approximated adder is also ineffective, due to the structure of the logic blocks.

These conclusions motivate for introducing new specialized \ac{dsp} blocks in FPGAs, implementing large adders fully in hardware.

\section{Acknowledgment}
This work was funded by the french ministry of higher education (MESR) and the LabEx IMobS$^3$ program at Institut Pascal (UMR 6602). We thank them and all the collaborators for their support to this research.

\bibliographystyle{ACM-Reference-Format}
\bibliography{Mendeley}

\end{document}